\documentclass[a4paper,11pt]{article}
\usepackage{pos}

\title{Spectrum of the SU(2) scalar-fermion-gauge system under the influence of the Brout-Englert-Higgs effect}
\ShortTitle{Spectrum of the SU(2) scalar-fermion-gauge system}

\author*{Georg Wieland}
\author{Axel Maas}

\affiliation{Institute of Physics, NAWI Graz, University of Graz,\\
  Universit{\"a}tsplatz 5, 8010 Graz, Austria}

\emailAdd{georg.wieland@uni-graz.at}
\emailAdd{axel.maas@uni-graz.at}

\abstract{Gauge invariance requires physical states to be composite, even in the weak sector of the Standard Model (SM). The Fr{\"o}hlich-Morchio-Strocchi (FMS) mechanism resolves this subtlety and predicts additional Higgs contributions in SM processes. While this has been supported by theoretical investigations in the bosonic sector, its impact on fermionic observables remains largely unexplored. We use non-perturbative lattice techniques within a gauge-invariant framework to study a proxy theory of the weak sector with dynamical fermions. We determine the physical spectrum of the theory and interpret the results in the context of the FMS mechanism. Additionally, we identify suitable simulation points for a scattering analysis as a first step toward cross-sections relevant to (future lepton) colliders.}

\FullConference{The European Physical Society Conference on High Energy Physics (EPS-HEP2025)\\
7-11 July 2025\\
Marseille, France\\}

\begin{document}
\maketitle

\section{Motivation}
\label{sec:intro}

Gauge theories and gauge-dependent degrees of freedom have proven to be valuable tools for describing the Standard Model (SM). However, introducing gauge fields as tools requires all observables, and consequently the spectrum of a gauge theory, to be gauge-invariant. For the weak part of the SM, this implies that the W/Z bosons, the Higgs field, and all left-handed fermions must be described by gauge-invariant composite objects and cannot be directly identified with elementary degrees of freedom. Despite this, the standard perturbative approach to weak physics, which uses elementary states as observables, reproduces experimental results with remarkable accuracy~\cite{Bohm:2001yx, ParticleDataGroup:2022pth}.

This conspicuous coincidence can be explained by the Fr{\"o}hlich-Morchio-Strocchi (FMS) mechanism~\cite{Frohlich:1980gj, Frohlich:1981yi}. Within the SM, the physical spectrum is in a one-to-one correspondence with the spectrum of elementary states. In other words, for each state charged under the weak gauge group, there exists a corresponding state charged under a global symmetry group. Consequently, asymptotic states can be represented by bound states carrying the same global quantum numbers as the elementary fields~\cite{Frohlich:1980gj, Frohlich:1981yi, Banks:1979fi}. This one-to-one mapping has been confirmed on the lattice for the W/Z-Higgs system~\cite{Shrock:1986fg, Maas:2013aia}, including quenched leptons~\cite{Afferrante:2020fhd}, and via continuum approaches~\cite{Maas:2020kda, Egger:2017tkd, Fernbach:2020tpa}. For a review, see~\cite{Maas:2017wzi}.

For the gauge-scalar part of the weak SM, genuine non-perturbative effects and therefore deviations from the standard perturbative treatment have been observed on the lattice~\cite{Maas:2018ska, Jenny:2022atm} and in analytical calculations~\cite{Dudal:2020uwb, Maas:2020kda, Egger:2017tkd, Maas:2022gdb}. However, the appearance of additional Higgs contributions in processes including fermions requires more attention -- particularly as these effects could be within reach of current and future experiments~\cite{Fernbach:2020tpa}.

A reliable non-perturbative, gauge-invariant treatment of a theory that combines gauge bosons, scalars, and fermions is an important endeavor in its own right. It is required not only for a manifestly gauge-invariant description of the SM, but also for beyond the SM scenarios that may not be adequately described by conventional perturbation theory~\cite{Maas:2015gma}. Lattice simulations allow us to combine all three types of particles and investigate the theory without gauge fixing, while also including all non-perturbative dynamics.

The following pages report on our progress in simulating Brout-Englert-Higgs (BEH) physics with dynamical fermions on the lattice. The theoretical background and the lattice setup are presented in Section~\ref{sec:background}, preliminary results are discussed in Section~\ref{sec:results}, and an outlook on ongoing and future work is provided in Section~\ref{sec:outlook}.

\section{Theoretical framework and lattice setup}
\label{sec:background}

\subsection{FMS mechanism}
\label{sec:fms}

The weak sector of the SM is accurately described by an SU(2) gauge theory coupled to a scalar field in the fundamental representation with an active BEH effect. In addition to the local gauge symmetry, a global one -- the so-called custodial symmetry -- remains as a residual symmetry of the gauging procedure. This residual symmetry is also SU(2), which allows for the one-to-one correspondence between physical and elementary states predicted by the FMS mechanism.

The standard textbook approach to the BEH effect proceeds by analyzing the quartic potential in the Lagrangian and expanding the scalar field around its minimum. This procedure spontaneously breaks the gauge symmetry and generates massive gauge bosons~\cite{Bohm:2001yx}. However, this approach relies on a gauge-dependent construction and is associated with several conceptual issues~\cite{Maas:2017wzi}. First, Elitzur's theorem forbids spontaneous gauge symmetry breaking. Second, the Gribov-Singer ambiguity makes gauge fixing non-unique beyond perturbation theory. Third, quantities such as the vacuum expectation value depend on the gauge choice.

A gauge-invariant reformulation based on the FMS mechanism addresses these issues by constructing composite operators with the same global quantum numbers as the corresponding elementary fields, i.e., $\phi^\dagger\phi$ (Higgs), $\phi^\dagger D_\mu \phi$ (W/Z bosons), and $\phi^\dagger\psi$ (left-handed fermions), where $\phi$, $W_\mu$\footnote{The gauge boson fields $W_\mu$ are understood to enter the covariant derivative $D_\mu = \partial_\mu + ig W_\mu$.}, and $\psi$ are the elementary degrees of freedom. Expanding the correlation functions of the composite states in terms of the vacuum expectation value in a suitable gauge maps the propagators of the composite states to the propagators of the elementary states at leading order~\cite{Frohlich:1980gj, Frohlich:1981yi}. Consequently, the physical states have the same masses as their elementary counterparts.

\subsection{SU(2) scalar-fermion-gauge system}
\label{sec:system}

Lattice studies of the weak sector of the SM are currently not possible because of two significant problems. First, the large scale separation between SM leptons and the Higgs makes computations prohibitively expensive for leptons with masses below a few tens of GeV. Second, and even more problematic, the chiral nature of the theory cannot be satisfactorily simulated on the lattice. Nevertheless, qualitative questions can still be answered using vectorial leptons~\cite{Afferrante:2020fhd}.

Therefore, we investigate an SU(2) Yang-Mills theory coupled to a gauged scalar doublet, representing the weak subsector of the SM, together with two mass-degenerate generations of vectorial leptons. Each lepton generation includes one weakly charged doublet and one ungauged doublet with an additional flavor symmetry~\cite{Afferrante:2020fhd}. This allows for adding gauge-invariant Yukawa terms with the same structure as in the SM. However, for our first simulations, we assume vanishing Yukawa couplings, which effectively decouples the ungauged flavors.

The gauge-invariant bound states predicted by the FMS mechanism in this theory are the scalar singlet $\mathcal{O}_{0^+}$, the vector triplet $\mathcal{O}_{1^-}^{\mathbf{a}, \mu}$, and the fermion bound state $\mathcal{O}_{\Psi}^{\mathbf{i},g}$. Here, $\mathbf{a}$ and $\mathbf{i}$ denote custodial indices, while $g$ labels the generation. $\mathcal{O}_{0^+}$ transforms as a singlet under the custodial group, whereas $\mathcal{O}_{1^-}^{\mathbf{a}, \mu}$ forms a triplet. For vanishing Yukawa couplings, the fermion operator transforms as a doublet under both the custodial and the global generation symmetry.

\subsection{Lattice details}
\label{sec:lattice}

To study the theory in a fully gauge-invariant and non-perturbative way, we discretize the underlying action on a four-dimensional Euclidean lattice with lattice spacing $a$ and lattice size~$L$ using standard techniques. To investigate the spectrum, we construct and measure lattice versions of the composite states introduced in the previous section (see~\cite{Evertz:1985fc, Afferrante:2020fhd, Jenny:2022atm} for details). Additionally, we introduce the operator $\mathcal{O}_L^g$ in the leptonium channel. This operator describes a gauge-invariant pseudo-scalar composite state of two leptons, analogous to the pion in Quantum Chromodynamics (QCD), and is a triplet under the global generation symmetry. We use standard techniques to construct the corresponding correlators and extract their masses. We simulate the theory using the publicly available HiRep code~\cite{DelDebbio:2008zf, Drach:2025eyg} with additional support for scalar fields~\cite{Hansen:2017mrt}.

For a first exploration of the theory, we introduce the fermions by gradually reducing their mass. This is accomplished by decreasing the bare fermion mass parameter $m_F$. In practice, we also tune the value of $m_F$ to identify specific points in parameter space, allowing us to explore various decay channels in a subsequent scattering analysis. The corresponding parameters are listed in Table~\ref{tab:param}. All simulations are performed on a cubic lattice with $L=16$.

\renewcommand{\arraystretch}{1.3}
\begin{table}
\centering
\begin{tabular}{c | c | c | c c c c c }
 $\beta$ & $m_S^2$ & $\lambda_S$ & & & $m_F$ & & \\
 \hline
  $8.0$ & $-0.4168$ & $0.1923$ & $0.3103$ & $0.06174$ & $-0.009577$ & $-0.02544$ & $-0.06299$ \\
\end{tabular}
\caption{Lattice parameters used in our simulations. The parameters include the gauge coupling $\beta$, the squared scalar mass $m_S^2$, the scalar quartic coupling $\lambda_S$, and the bare fermion mass $m_F$. The parameters of the gauge-scalar subsystem are taken from~\cite{Wurtz:2013ova}.}
\label{tab:param}
\end{table}
\renewcommand{\arraystretch}{1}

\section{Preliminary results}
\label{sec:results}

\begin{figure}
  \centering
  \includegraphics[scale=0.75]{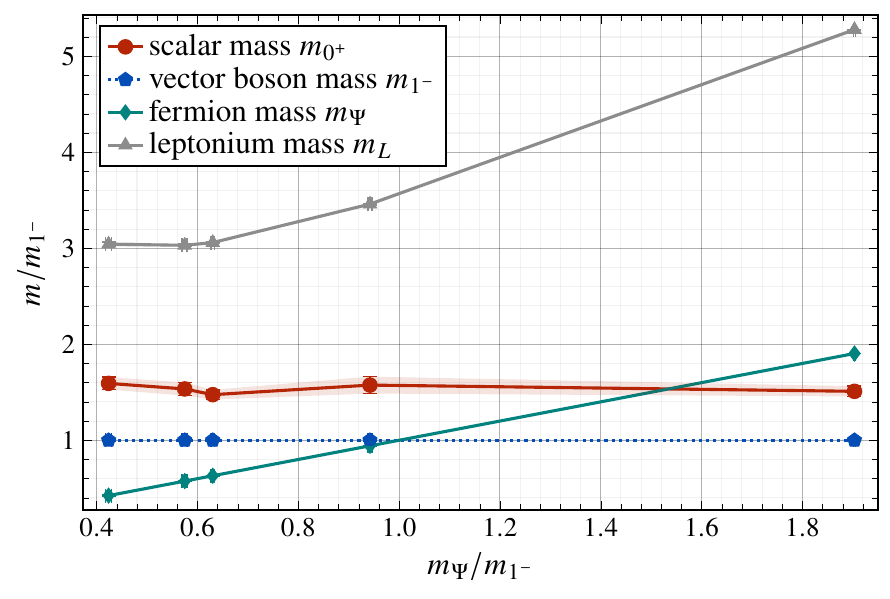}
  \caption{Dependence of the scalar, vector boson, fermion, and leptonium masses on the fermion mass, computed for the parameters listed in Table~\ref{tab:param}. All masses are extracted from correlators constructed from gauge-invariant composite operators. Error bars are mostly smaller than the symbol size.}
  \label{fig:k-dep}
\end{figure}

The results of our lattice analysis of an SM-proxy theory with vectorial fermions are shown in Figure~\ref{fig:k-dep}, which depicts the dependence of the bound-state masses on the constituent fermion mass. All quantities are expressed as dimensionless ratios relative to the vector boson mass. In total, ground-state energy levels in four channels, corresponding to masses of stable states, are considered: the scalar ($m_{0^+}$), the vector boson ($m_{1^-}$), the fermion ($m_{\Psi}$), and the leptonium channel ($m_L$). The results are assumed to be in the Higgs-like domain of the phase diagram, as $m_{0^+}>m_{1^-}$. This assumption is supported by simulations of the gauge-scalar subsystem, both without and with heavy fermions, as well as by comparison with data presented in~\cite{Wurtz:2013ova}. Furthermore, we do not observe any hints of chiral symmetry breaking in the leptonium channel, which suggests that the underlying dynamics are not QCD-like but instead resemble a BEH-like realization of the theory.

As shown in Figure~\ref{fig:k-dep}, we find a stable Higgs state over a wide range of fermion masses. The results also show an SM-like ratio for $m_{0^+}/m_{1^-}$, which is consistent with simulations of the gauge-scalar subsystem~\cite{Wurtz:2013ova}. This ratio remains constant as the fermion mass decreases, i.e., the fermion mass can be varied from above $m_{0^+}$ to below $m_{1^-}$ while keeping the mass ratio $m_{0^+}/m_{1^-}$ constant within errors. Our results reveal that the gauge-scalar dynamics of the studied system remain unchanged upon the inclusion of fermions, which indicates that the behavior of the system can be fully controlled by adjusting the fermion mass. This feature appears to be generic in the Higgs-like domain, highlighting the robustness of the observed dynamics~\cite{Maas:2025abc}.

Comparing our findings with expectations from perturbation theory, we observe that an interpretation of our results within the FMS framework provides a consistent description. The physical fermion exhibits a significant mass defect, particularly for small $m_\Psi/m_{1^-}$, with its mass dropping below $m_{0^+}$. Moreover, the leptonium channel is consistent with a scattering state of two physical leptons. This indicates additional Higgs contributions that perturbation theory cannot account for but are naturally accommodated by FMS predictions. For very small $m_\Psi/m_{1^-}$, overlap effects partially obscure this picture.

\section{Outlook}
\label{sec:outlook}

Using lattice gauge theory, we have demonstrated our ability to determine the physical spectrum of an SM-proxy theory, including the Higgs, W/Z bosons, and leptons, in a manifestly gauge-invariant approach. Based on a comparison with perturbative expectations, our findings appear to favor an interpretation of weak physics in terms of the FMS mechanism.

The next step in this ongoing investigation is to obtain additional higher energy levels. Acquiring these is essential for performing a L{\"u}scher analysis~\cite{Luscher:1986pf}, which establishes a connection between the volume-dependent energy levels and the infinite-volume phase-shifts. From there, we are particularly interested in calculating cross-sections, along the lines of~\cite{Jenny:2022atm} for the quenched case, as these quantities allow for a direct comparison with experimental data.

We aim to elucidate qualitative differences between a conventional perturbative and a fully gauge-invariant formulation of weak physics. Our goal is to establish a direct connection with experimental observations by analyzing the implications of a gauge-invariant description of scattering processes within the SM. This approach ensures that experimental signals from current and future lepton colliders are understood within the complete context of the SM, rather than being misidentified as evidence for new physics.

\acknowledgments

We would like to thank Yannick Dengler, Sofie Martins, and Fabian Zierler for helpful discussions. GW is supported by the Austrian Science Fund FWF, grant PAT6443923. The computational results have been achieved using the Austrian Scientific Computing (ASC5) infrastructure and the Graz Scientific Cluster (GSC1).

\bibliographystyle{JHEP}
\bibliography{main.bib}

\providecommand{\href}[2]{#2}\begingroup\raggedright\begin{thebibliography}{10}

\bibitem{Bohm:2001yx}
M.~B{\"o}hm, A.~Denner and H.~Joos, \emph{{Gauge theories of the strong and electroweak interaction}}, Teubner, Stuttgart (2001), \href{https://doi.org/10.1007/978-3-322-80160-9}{10.1007/978-3-322-80160-9}.

\bibitem{ParticleDataGroup:2022pth}
{\scshape Particle Data Group} collaboration, \emph{{Review of Particle Physics}}, \href{https://doi.org/10.1093/ptep/ptac097}{\emph{PTEP} {\bfseries 2022} (2022) 083C01}.

\bibitem{Frohlich:1980gj}
J.~Fr{\"o}hlich, G.~Morchio and F.~Strocchi, \emph{{Higgs phenomenon without a symmetry breaking order parameter}}, \href{https://doi.org/10.1016/0370-2693(80)90594-8}{\emph{Phys. Lett. B} {\bfseries 97} (1980) 249}.

\bibitem{Frohlich:1981yi}
J.~Fr{\"o}hlich, G.~Morchio and F.~Strocchi, \emph{{Higgs phenomenon without symmetry breaking order parameter}}, \href{https://doi.org/10.1016/0550-3213(81)90448-X}{\emph{Nucl. Phys. B} {\bfseries 190} (1981) 553}.

\bibitem{Banks:1979fi}
T.~Banks and E.~Rabinovici, \emph{{Finite Temperature Behavior of the Lattice Abelian Higgs Model}}, \href{https://doi.org/10.1016/0550-3213(79)90064-6}{\emph{Nucl. Phys. B} {\bfseries 160} (1979) 349}.

\bibitem{Shrock:1986fg}
R.E.~Shrock, \emph{{Lattice Gauge Higgs Theories With Local X Global Symmetry Groups: General Properties and Exact Solutions}}, \href{https://doi.org/10.1103/PhysRevLett.56.2124}{\emph{Phys. Rev. Lett.} {\bfseries 56} (1986) 2124}.

\bibitem{Maas:2013aia}
A.~Maas and T.~Mufti, \emph{{Two- and three-point functions in Landau gauge Yang-Mills-Higgs theory}}, \href{https://doi.org/10.1007/JHEP04(2014)006}{\emph{JHEP} {\bfseries 04} (2014) 006} [\href{https://arxiv.org/abs/1312.4873}{{\ttfamily 1312.4873}}].

\bibitem{Afferrante:2020fhd}
V.~Afferrante, A.~Maas, R.~Sondenheimer and P.~T{\"o}rek, \emph{{Testing the mechanism of lepton compositeness}}, \href{https://doi.org/10.21468/SciPostPhys.10.3.062}{\emph{SciPost Phys.} {\bfseries 10} (2021) 062} [\href{https://arxiv.org/abs/2011.02301}{{\ttfamily 2011.02301}}].

\bibitem{Maas:2020kda}
A.~Maas and R.~Sondenheimer, \emph{{Gauge-invariant description of the Higgs resonance and its phenomenological implications}}, \href{https://doi.org/10.1103/PhysRevD.102.113001}{\emph{Phys. Rev. D} {\bfseries 102} (2020) 113001} [\href{https://arxiv.org/abs/2009.06671}{{\ttfamily 2009.06671}}].

\bibitem{Egger:2017tkd}
L.~Egger, A.~Maas and R.~Sondenheimer, \emph{{Pair production processes and flavor in gauge-invariant perturbation theory}}, \href{https://doi.org/10.1142/S0217732317502121}{\emph{Mod. Phys. Lett. A} {\bfseries 32} (2017) 1750212} [\href{https://arxiv.org/abs/1701.02881}{{\ttfamily 1701.02881}}].

\bibitem{Fernbach:2020tpa}
S.~Fernbach, L.~Lechner, A.~Maas, S.~Pl{\"a}tzer and R.~Sch{\"o}fbeck, \emph{{Constraining the Higgs boson valence contribution in the proton}}, \href{https://doi.org/10.1103/PhysRevD.101.114018}{\emph{Phys. Rev. D} {\bfseries 101} (2020) 114018} [\href{https://arxiv.org/abs/2002.01688}{{\ttfamily 2002.01688}}].

\bibitem{Maas:2017wzi}
A.~Maas, \emph{{Brout-Englert-Higgs physics: From foundations to phenomenology}}, \href{https://doi.org/10.1016/j.ppnp.2019.02.003}{\emph{Prog. Part. Nucl. Phys.} {\bfseries 106} (2019) 132} [\href{https://arxiv.org/abs/1712.04721}{{\ttfamily 1712.04721}}].

\bibitem{Maas:2018ska}
A.~Maas, S.~Raubitzek and P.~T{\"o}rek, \emph{{Exploratory study of the off-shell properties of the weak vector bosons}}, \href{https://doi.org/10.1103/PhysRevD.99.074509}{\emph{Phys. Rev. D} {\bfseries 99} (2019) 074509} [\href{https://arxiv.org/abs/1811.03395}{{\ttfamily 1811.03395}}].

\bibitem{Jenny:2022atm}
P.~Jenny, A.~Maas and B.~Riederer, \emph{{Vector boson scattering from the lattice}}, \href{https://doi.org/10.1103/PhysRevD.105.114513}{\emph{Phys. Rev. D} {\bfseries 105} (2022) 114513} [\href{https://arxiv.org/abs/2204.02756}{{\ttfamily 2204.02756}}].

\bibitem{Dudal:2020uwb}
D.~Dudal, D.M.~van Egmond, M.S.~Guimaraes, L.F.~Palhares, G.~Peruzzo and S.P.~Sorella, \emph{{Spectral properties of local gauge invariant composite operators in the $SU(2)$ Yang{\textendash}Mills{\textendash}Higgs model}}, \href{https://doi.org/10.1140/epjc/s10052-021-09008-9}{\emph{Eur. Phys. J. C} {\bfseries 81} (2021) 222} [\href{https://arxiv.org/abs/2008.07813}{{\ttfamily 2008.07813}}].

\bibitem{Maas:2022gdb}
A.~Maas and F.~Reiner, \emph{{Restoring the Bloch-Nordsieck theorem in the electroweak sector of the standard model}}, \href{https://doi.org/10.1103/PhysRevD.108.013001}{\emph{Phys. Rev. D} {\bfseries 108} (2023) 013001} [\href{https://arxiv.org/abs/2212.08470}{{\ttfamily 2212.08470}}].

\bibitem{Maas:2015gma}
A.~Maas, \emph{{Field theory as a tool to constrain new physics models}}, \href{https://doi.org/10.1142/S0217732315501357}{\emph{Mod. Phys. Lett. A} {\bfseries 30} (2015) 1550135} [\href{https://arxiv.org/abs/1502.02421}{{\ttfamily 1502.02421}}].

\bibitem{Evertz:1985fc}
H.G.~Evertz, J.~Jersak, C.B.~Lang and T.~Neuhaus, \emph{{SU(2) Higgs Boson and Vector Boson Masses on the Lattice}}, \href{https://doi.org/10.1016/0370-2693(86)91547-9}{\emph{Phys. Lett. B} {\bfseries 171} (1986) 271}.

\bibitem{DelDebbio:2008zf}
L.~Del~Debbio, A.~Patella and C.~Pica, \emph{{Higher representations on the lattice: Numerical simulations. SU(2) with adjoint fermions}}, \href{https://doi.org/10.1103/PhysRevD.81.094503}{\emph{Phys. Rev. D} {\bfseries 81} (2010) 094503} [\href{https://arxiv.org/abs/0805.2058}{{\ttfamily 0805.2058}}].

\bibitem{Drach:2025eyg}
V.~Drach, S.~Martins, C.~Pica and A.~Rago, \emph{{High-Performance Simulations of Higher Representations of Wilson Fermions}},  \href{https://arxiv.org/abs/2503.06721}{{\ttfamily 2503.06721}}.

\bibitem{Hansen:2017mrt}
M.~Hansen, T.~Janowski, C.~Pica and A.~Toniato, \emph{{SU(2) with fundamental fermions and scalars}}, \href{https://doi.org/10.1051/epjconf/201817508010}{\emph{EPJ Web Conf.} {\bfseries 175} (2018) 08010} [\href{https://arxiv.org/abs/1710.10831}{{\ttfamily 1710.10831}}].

\bibitem{Wurtz:2013ova}
M.~Wurtz and R.~Lewis, \emph{{Higgs and W boson spectrum from lattice simulations}}, \href{https://doi.org/10.1103/PhysRevD.88.054510}{\emph{Phys. Rev. D} {\bfseries 88} (2013) 054510} [\href{https://arxiv.org/abs/1307.1492}{{\ttfamily 1307.1492}}].

\bibitem{Maas:2025abc}
A.~Maas, S.~Martins and G.~Wieland, \emph{{Lattice analysis of the SU(2) scalar-fermion-gauge system}}, {\emph{unpublished}}.

\bibitem{Luscher:1986pf}
M.~L{\"u}scher, \emph{{Volume Dependence of the Energy Spectrum in Massive Quantum Field Theories. 2. Scattering States}}, \href{https://doi.org/10.1007/BF01211097}{\emph{Commun. Math. Phys.} {\bfseries 105} (1986) 153}.

\end{thebibliography}\endgroup

\end{document}